\documentclass[a4paper,12pt,reqno]{amsart}

\usepackage{amssymb,amsmath,amsfonts}

\usepackage[all]{xy}
\input xy

\theoremstyle{definition}
\newtheorem{definition}{Definition}

\newcounter{example}
\newenvironment{example}[1]{%
  \vspace{\medskipamount}
  \addtocounter{example}{1}%
  \noindent\textbf{Example \arabic{example}. } \textit{#1}\par\noindent}{%
  \ensuremath{\Box}\par}

\begin{document}
\title[Syntax Diagrams]{Syntax diagrams as a formalism for representation of syntactic relations of formal languages}
\author{Vladimir Lapshin}
\makeatletter
\newdimen\whitespaceamount
\settowidth\whitespaceamount{~}
\def\@@and{\hskip-\whitespaceamount,~}
\makeatother

\sloppy

\begin{abstract}
The new approach to representation of syntax of formal languages~-- a formalism of syntax diagrams is offered. Syntax diagrams look a convenient language for the description of syntactic relations in the languages having nonlinear representation of texts, for example, for representation of syntax lows of the language of structural chemical formulas. The formalism of neighbourhood grammar is used to describe the set of correct syntax constructs. The neighbourhood the grammar consists of a set of families of "neighbourhoods"~-- the diagrams defined for each symbol of the language's alphabet. The syntax diagram is correct if each symbol is included into this diagram together with some neighbourhood. In other words, correct diagrams are needed to be covered by elements of the neighbourhood grammar. Thus, the grammar of formal language can be represented as system of the covers defined for each correct syntax diagram.
\end{abstract}
\maketitle

\section{The work's motivation}

The idea of representation of syntax relations of a formal language by means of the definition of families of language symbols' neighbourhoods belongs to Soviet mathematician J. Shreider (\cite {Shreider67}). The neighbourhood of a symbol here is understood as any chain of symbols containing this symbol. The chain is in the source language if, and only if each symbol belongs  to this chain together with some it's neighbourhood. Such the system of neighbourhoods has been named by Shreider as a neighbourhood grammar. Let consider a concrete example. Let $L$ be a formal language with the alphabet $A = \{a, b \}$ and chains of language $L$ are the sequences of alternating symbols $a$ and $b$, where first and last symbols must be $a$. In other words, chains of language $L$ are chains of a kind $aba$, $ababa$, $abababa$, etc. Let define the neighbourhood grammar for this language by enumerating a finite system of neighbourhoods for each symbol of the alphabet $A$. Let consider the symbol $a$ and places in the language's chains where it is occurred. This symbol necessarily appears in the beginning and the end of any chain of formal language $L$. To accent this fact, enter an additional pseudo-symbol $\#$ which will signal about the beginning and the end of a chain. Thus, there are two neighbourhoods of a symbol $a$: a neighbourhood $\#ab $ and a neighbourhood $ba\#$. Except for the above-stated cases, the symbol $a$ can be between two symbols $b$. Add for this case a neighbourhood $bab$ a symbol $a$. For a symbol $b$ enough a unique neighbourhood~-- chains $aba$. So, any chain of language $L$ becomes covered by the neighbourhoods specified above. It is easy to prove the contrary: any chain which becomes covered by the system of neighbourhoods defined above belongs to language $L$. Languages for which it is possible to define a neighbourhood grammar in sense of Shreider are named as Shreider's ones. Shreider's languages are simple enough in sense of expression of syntactic relations. The unique type of syntactic laws which can be expressed by neighbourhood grammars is the relation "to be close to". In Chomsky's hierarchy Shreider's languages represent own subset of linear languages. In other words, neighbourhood grammars as they be formulated by Shreider cannot be used to define the overwhelming majority of languages.

The idea of neighbourhood grammar developed in works of the Soviet mathematicians of V.Borschev and M.Homyakov (\cite {Borschev70}, \cite {Borschev73}). They have suggested to expand a traditional sight at formal language as on a set of chains defined on some alphabet. In Borschev and Homyakov's works a neighbourhood grammar was used to define not chains of symbols, but wider concept of texts. Texts could represent everything: chemical formulas, graphs and etc. In particular, the neighbourhood interpretation of context-free languages has been offered. As it is known, each chain belongs to context-free language has at least one derivation tree. Such tree has the top signed by an initial non-terminal symbol of the context-free grammar for the given language, internal units are signed by non-terminal symbols, and sheet units signed by terminals of the given grammar. The idea was to define the set of correctly constructed derivation trees of the given context-free language by a systems of neighbourhoods, defined for each symbol of the language (nonterminal and terminal one). The neighbourhood is understood here as some subtree containing the dedicated symbol~-- the center of the neighbourhood. Borschev and Homyakov found that such the neighbourhood grammar can be defined for each context-free language. The neighbourhoods of the grammar are either a bush consisting of the one level tree, where the center of the neighbourhood is the top nonterminal of a bush, or one knot tree consisting of a single terminal symbol. The grammar's nonterminals have the first type of neighbourhoods (bushes) and single vertex trees are the neighbourhoods of the terminals. Borschev and Homyakov used the formalism of model theory to define neighbourhood grammars. There were two sorts of axioms. The first one was used to describe the set of possible trees that can be built on the given alphabet. And the second sort of axioms (neighbourhoods) was used to select from the set of all trees, defined by the first sort axioms, the set of correct derivation trees. An each derivation tree had been defined as the model of the theory. It is important for our purpose that syntax relations of the language obviously hidden in the grammar's rules here are visualized by derivation trees description. For instance, the neighbourhood grammar were successfully applied to define the language of structural chemical formulas, that describe complex organic compounds (\cite{Pantuhina72}).

The representation of the language as the set of symbols, connected by complex syntax relations, may be generalized by using of the formalism of \textit{syntax diagram}. The syntax diagram here is understanding as multigraph (i.e. a graph which nodes are connected more then one rib), which nodes are signed by symbols from some alphabet. The nodes describe some atomic entities (for example, symbols or groups of symbols) and ribs describe the syntax relations between them. There are many objects that could be thought as syntax diagrams. That are derivation trees, structural chemical formulas and formal language's chains, that are represented as graphs where the rib goes from the symbol of a chain to the previous one. The approach to definition of formal language's symbol chains with description of local syntax structures for an each symbol of the language, which was suggested by Shreider, may be naturally generalized to define the set of correct syntax diagrams. It is needed only to define the finite family of syntax diagrams for each symbol of the alphabet. An each diagram from the family for the symbol must contain at least on node signed by this symbol and one of such nodes became the center of the neighbourhood defined by this syntax diagram. The correct syntax diagram is defined as the syntax diagram that has each node together with some its neighbourhood as a subdiagram of this one. Further we shall define the notion of "subdiagram" more closely.

\section{Syntax diagrams}

The syntax diagram will be defined as a connected multigraph, which ribs can belongs to different sorts. The using of more than one sort to name the ribs is very comfortable to express different syntax relations in the language. For instance, to describe derivation trees as the syntax diagrams it is naturally to use two sorts of relations and, accordingly, ribs. The first sort express the relation between nodes on the same level of the tree, which can be named as "to be left on". And the second sort express the relation between parent and child nodes in the tree. The multigraph of a syntax diagram could be directed or not, it depends on the syntax of the defining language. There can be more then one rib between two nodes and every rib can belongs to different sorts. It is not permitted to define ribs connected the same node, this does not any sense in the syntax description because the nodes, signed the same symbol, always describe the same atomic entity and the rib, connected such the nodes, naturally express any syntax relation of the entity to the same one. Thus, it is not clear what kind of syntax property can be expressed by a loop in the multigraph.

\begin{definition}
The many sorted multigraph can be defined as quadruple $\Gamma=\{V,R,S,f\}$ where $V=\{v_1,v_2,\ldots,v_m\}$~-- finite set of nodes, $R$~-- finite set of ribs (pairs from set $V$ where pairs $(v,v):v \in v$ are not permitted), $S=\{S^1,S^2,\ldots,S^k\}$--- finite set of sorts and $f:R \rightarrow S$-- mapping of sortification, that gives some sort to an each rib. Multigraph $\Gamma$ is named as "directed" one if pairs from $R$ are ordered, in the contrary case the graph is named as "undirected". The path in the multigraph $\Gamma=\{V,R,S,f\}$ is the sequence $P=\{(v_{i_1},v_{i_2}),(v_{i_2},v_{i_3}),\ldots,(v_{i_{n-1}},v_{i_n})\}$ pairs of nodes where each pair belongs to $R$ and the second node in the current pair is the same as the first node in the next one. The many sorted multigraph $\Gamma=\{V,R,S,f\}$ is connected if for any pair of nodes $v_i,v_j \in V$ exists a path $P=\{(v_{i_1},v_{i_2}),(v_{i_2},v_{i_3}),\ldots,(v_{i_{n-1}},v_{i_n})\}$, such that $v_i = v_{i_1}$ and $v_j = v_{i_n}$.
\end{definition}

Note that a path in the many sorted multigraph can connect nodes via ribs of different sorts. Also, the ribs' direction is not taken in account in the path's definition.

\begin{definition}
Let $A=\{a_1,a_2,\ldots,a_n\}$ be an alphabet (finite set of symbols), $\Gamma=\{V,R,S,f\}$~-- many sorted connected multigraph and $F:V \rightarrow A$~-- naming mapping, which maps each node of $\Gamma$ to some symbol of $A$. The triplet $D=\{A,\Gamma,F\}$ will be named further as a syntax diagram.
\end{definition}

It is convenient to use syntax diagrams to describe texts that have nonlinear representation. It is, for example, structural chemical formulas. An each full structural formula is exactly the syntax diagram based on the undirected multigraph, which nodes are signed by chemical elements names and ribs represent covalent bonds. There is only the single sort for all ribs of the syntax diagram. For instance, molecule ${\sf H_2O}$ is represented by syntax diagram $\xymatrix@1{H \ar@{-}[r] & O \ar@{-}[r] & H}$. The method can be also applied to describe texts that have linear representation. Each such representation is the syntax diagram based on directed graph, where each rib connects the next node with the previous one and nodes are signed by symbols of the alphabet. An each node of such the diagram has not more then one incoming rib and one outcoming rib and there is the single node (the first node of the chain), which has only one rib~-- incoming one, and the single node, which has only the single outcoming rib (the last node of the chain). So, the ribs here just represent the linear order on the text. For example, if $abcaabc$ be a symbol chain, then the syntax diagram for it be $$\xymatrix{a & b \ar@{>}[l] & c \ar@{>}[l] & a \ar@{>}[l] & a \ar@{>}[l] & b \ar@{>}[l] & c \ar@{>}[l]}$$

Often it is convenient to think not about all set syntax diagrams on the given alphabet and sorts, but select some subset of this one. It has been done above when the method of representation of linear texts as syntax diagrams were discussed. Such the selection cab be done by using some \textit{restrictions} on the structure of diagrams' multigraphs. Again, to describe symbol chains as the syntax diagrams it is convenient to put some restrictions on nodes and ribs of such the diagrams as is has been done above. This method will always be used further elsewhere: when the language of syntax diagrams will be defined, some restrictions should be defined as well. The restrictions can be defined by enumerating an alphabet, nodes signed by symbols of the alphabet, sorts of ribs and definitions saying which nodes can be connected by ribs of the given sorts. The restrictions describe the syntax of the language globally, by applying to each correct syntactic construction, as opposing there is local definitions of the syntax, applying to each symbol of the language, these what named as neighbourhoods.

The neighbourhood grammars make possible to select correct syntax constructs from all set of syntax diagrams satisfying the given restrictions. Thus, a neighbourhood grammar describes the language of syntax diagrams by defining:
\begin{enumerate}
	\item Globally: the finite set of restrictions that each syntax diagram should satisfy with.
	\item Locally: the finite family of neighbourhood diagrams defined for each symbol of the language.
\end{enumerate}

\section{Neighbourhood grammars and syntax covers}

Firstly, precisely define what is a syntax subdiagram. This notion can be described in terms of mappings between nodes and ribs of diagram and subdiagram, which save the sorts of ribs and naming of nodes. Let $h(v',v'')$ be the set of ribs that connect nodes $v'$ and $v''$, and $h_{s'}(v',v'')$~-- the set of ribs having the sort $s'$ and connecting nodes $v'$ and $v''$.

\begin{definition}
Let $A$ be an alphabet and $D_1=\{A,{\Gamma}_1,F_1\}$, $D_2=\{A,{\Gamma}_2,F_2\}$ be syntax diagrams. The triple $s=(s_V,s_R,s_S)$ of injective mappings $s_V:V_1 \rightarrow V_2$, $s_R:R_1 \rightarrow R_2$ and $s_S:S_1 \rightarrow S_2$ will be generally named as inclusion mapping, and pair $(D_1,s)$~-- syntax subdiagram of syntax diagram $D_2$, if mappings $s=(s_V,s_R,s_S)$ satisfy following conditions:
\begin{enumerate}
   \item $F_1(v)=F_2(s_V(v))$ for each $v \in V_1$.
   \item $s_S(f_1(v',v''))=f_2(s_R(v',v''))$ for each $(v',v'') \in R_1$.
   \item $s_R(v',v'')=(s_V(v'),s_V(v''))$ for each $(v',v'') \in R_1$.
\end{enumerate}
\end{definition}

The first two conditions just fix the fact that inclusion mapping $s=(s_V,s_R,s_S)$ should save the nodes' naming and ribs' sorting. The third condition correctly connects the naming and sorting mapping together to each ribs in the subdiagram maps to images of its nodes in the including diagram.

The example below demonstrates the multiformity of different inclusions a diagram to an another one. Let it given diagrams $D_1 = $ $\xymatrix@1{a & b \ar@{>}[l] \ar@{>}[r] & a}$ and $D_2 = $ $\xymatrix@1{a & b \ar@{>}[l]}$. The diagram $D_2$ is included to $D_1$ as two subdiagrams. The first one is defined by inclusion mapping, which maps node $a$ of $D_2$ to the first node $a$ of $D_1$. The second mapping maps, accordingly, node $a$ to the second node $a$ of $D_1$. Both inclusions map node $b$ of diagram $D_2$ by the only possible way to node $b$ of diagram $D_1$. This shows the possibility to have more then one inclusion mapping between two syntax diagrams. It is where syntax diagrams are differed from sets. Also, one can say about the diagram, which consists of only the single node or about the diagram, which does not contain any node~-- empty diagram. It is naturally think that the empty diagram is contained to any diagram from the given set.

The idea of Yuliy Shreider was about the syntactically correct symbol chains must be covered by neighbourhoods of their symbols. It should be defined the finite family of neighbourhoods (chains that contain the certain symbol) for each symbol of the alphabet and such the symbol must be selected in the chain. The neighbourhood grammar for the language of syntax diagrams will be defined by analogy with Shreider's idea for the language of symbol chains. but, firstly define what is the neighbourhood of a symbol in a grammar of syntax diagrams' language. 

\begin{definition}
The neighbourhood $D_a$ of the symbol $a \in A$ is the pair $(D_a,s_a)$, where $D_a$~-- the syntax diagram and $s_a : a \rightarrow D_a$~-- inclusion mapping of syntax diagram $a$, which contains only the single node signed by the symbol $a$, to the syntax diagram $D_a$. The node $s_a(a)$ will be named as the center of neighbourhood $D_a$.
\end{definition}

A it has been noted above, a syntax diagram is defined by three factors: an alphabet $A$, a finite set of sorts $S$ and restrictions $C$. So, it makes sense to say about the set of syntax diagram ${\bf D}=\{A,S,C\}$.

\begin{definition}
Let ${\bf D}=\{A,S,C\}$ be a set of syntax diagrams. The neighbourhood grammar $G$, which is defined on the set ${\bf D}$, is the finite family $G=\{G_a : a \in A, G_a \in {\bf D} \}$ of neighbourhoods defined for each symbol $a$ of alphabet $A$.
\end{definition}

Thus, a neighbourhood grammar defines the family of neighbourhoods $G_a$ for each symbol $a \in A$. And such the family $G$, defined on the diagrams' set ${\bf D}=\{A,S,C\}$, is named as a neighbourhood grammar on the set ${\bf D}$. But, the question is how a neighbourhood grammar allows to differ correct syntax diagrams on the set ${\bf D}$ from incorrect ones? To understand this it is needed to enter the notion of the \textit{star} of a syntax diagram in a node $v$.

\begin{definition}
Let $D$ be a syntax diagram and $z$~-- function, which maps each node $v$ of $D$ to set of ribs of diagram $D$ that connect $v$ to another node or connect another node to $v$. The star of the syntax diagram $D$ in the node $v$ is the set $z(D)(v)$.
\end{definition}

One can also say about the star of a neighbourhood. An each a neighbourhood has the center~-- the selected node. So, the star of a a neighbourhood is the star of a neighbourhood's diagram in the a neighbourhood's center. Now, it is time going to the definition of the syntax cover.

\begin{definition}
Let ${\bf D}=\{A,S,C\}$ be a set of syntax diagrams, $G=\{G_a : a \in A\}$~-- a neighbourhood grammar defined on set ${\bf D}$ and $D \in {\bf D}$~-- some syntax diagram having the set of nodes $V$. The family $G_D=\{D_v : v \in V, D_v \in G_{F(v)}\}$ of neighbourhoods will be named as the syntax cover of syntax diagram $D$, if following conditions are true:
\begin{enumerate}
	\item For each node $v$ of diagram $D$ the neighbourhood $D_v$ is subdiagram $(D_v,s^v)$ of diagram $D$.
	\item If $D_v$~-- the neighbourhood of node $v$ of diagram $D$, then $s^v_R(z(D_v)(v))=z(D)(v)$ should be true.
\end{enumerate}
\end{definition}

The diagram $D$ is a correct syntax diagram in the neighbourhood grammar $G$, if there some syntax cover of diagram $D$ by neighbourhoods that belong to $G$. Thus, the syntax cover of a diagram is some family of neighbourhoods, defined for each node of the diagram. And each such the neighbourhood contains the star of the diagram in this node. This gives the method, which makes it possible to differ correct syntax diagrams from the incorrect ones on the given diagrams set.

Sometime, when there are many correct syntax diagrams that differ only by names of nodes, it is convenient to use variables on the alphabet. The variable is the element of an extra alphabet, which has empty refinement with alphabet $A$. There is the partial function, which maps some symbols of alphabet $A$ to names of variables. This makes possible to use only the single diagram for description of many syntax diagrams that differ each other only in names of nodes. Obviously, this approach is used to define many structurally like neighbourhoods. If there is a neighbourhood, where some node is signed by a variable, this means that there exists a lot of neighbourhoods and each of them may be got from the neighbourhood with a variable by changing the variable name to a symbol, which is in the variable's symbols set. In the next section there will be the example of the language of structural chemical formulas where the the variable are used.

\section{Examples}

\begin{example}{Formal language $aba$, $ababa$... (Shreider's languages)}%
Let $L(\{a,b\})=\{aba,ababa,abababa\ldots\}$ be a formal language. The chains of the language $L$ can be represented as syntax diagrams on the alphabet $\{a,b\}$. The ribs of such the diagrams have only one sort and all graphs of the diagrams are directed. The restrictions applying to the diagrams are follows:
\begin{enumerate}
	\item For an each diagram, each node of this one, except two, has exactly one incoming rib and one outcoming rib.
	\item For an each diagram, there is exactly one node, which only one rib~-- incoming and exactly one node, which has only one rib~-- outcoming.
\end{enumerate}

So, each such the diagram is the representation of some chain on the alphabet $\{a,b\}$, and for each two symbols of the chain there is one rib, directed from the next symbol to the previous one. For instance, the chain $aba$ is represented by diagram $\xymatrix@1{a & b \ar@{>}[l] & a \ar@{>}[l]}$. Define the neighbourhood grammar to select from this set of diagrams the diagrams that represent the chains of the language $L$. Define the neighbourhoods of symbol $a$, this are $\xymatrix@1{a & b \ar@{>}[l]}$, $\xymatrix@1{b & a \ar@{>}[l]}$ and $\xymatrix@1{b & a \ar@{>}[l] & b \ar@{>}[l]}$ (here it not needs to select the center of the neighbourhood, there is only one node signed by symbol $a$ in each diagram). For symbol $b$ only one neighbourhood should be defined, it is $a \leftarrow b \leftarrow a$. It is not difficult to show that each element of the language $D(L)$ of syntax diagrams, represented chains of the language $L$, has the syntax cover consisting of defined above neighbourhoods. And it easy to prove the contrary proposition: an each syntax diagram, which covered by defined above neighbourhoods, represents some chain of the language $L$.

The formal language $L$ is an example of a Shreider's language. It's not difficult to see that for an any Shreider's language there may be found the neighbourhood grammar as it has been described in the example.
\end{example}

\begin{example}{Context-free languages}%
As it is known, each chain of context-free language has at least one derivation tree. The idea is to represent all such the trees as syntax diagrams and define the neighbourhood grammar to select from the set of all diagram trees syntax diagrams that represent exactly derivation trees. In that case the neighbourhood grammar will be equivalent to the given context-free generative grammar. So, let $G=\{N,T,R,S\}$ be some context-free grammar, where $N$ is alphabet of nonterminal symbols, $T$ is alphabet of terminal symbols, $R$~-- set of rules and $S$~-- start nonterminal. Define the set $\textbf{D}=\{A,S,C\}$ of syntax diagrams for the grammar $G$. The diagrams are directed graphs. Let $A=N \bigcup T$ and let the set of sorts consist of two elements: $S_L$ (signs ribs from the next node of the tree to the previous one on the same level) and $S_P$ (from the parent node to the child one). The restrictions are follows:
\begin{enumerate}
	\item For an each diagram, each node except only the single one, has exactly one incoming rib having the sort $S_P$.
	\item For an each diagram, there is exactly one node, which has not any incoming rib.
	\item For an each diagram, an each node signed by terminal symbol, may have not more then one outcoming rib and, if such the rib exists, this rib has signed as $S_L$.
	\item For an each diagram, an each node, except only one, has not more then one incoming rib signed by sort $S_L$ and not more then one outcoming rib signed by sort $S_L$.
\end{enumerate}
The only one difference from the trivial derivation tree is. On an each level of a tree the linear order of nodes is exactly noted by ribs of sort $S_L$. When people draw the trees on the paper such the order is shown by natural way and it not any needs to draw additional ribs. The neighbourhood grammar on the set $\textbf{D}$ can be defined basing on the context-free grammar $G$. for each symbol of alphabet $A=N \bigcup T$ define the family of its neighbourhoods as the set of rules where this symbol includes to. For example, if symbol $a$ is the terminal one, then the neighbourhoods are all the rules $A \rightarrow X_1 \ldots X_k a X_{k+1} \ldots X_n$ and if symbol $a$ is in the more then one place in the right part of the rule, there must be the special neighbourhood for this including. The same needs to be done for nonterminal symbols, taking in account in addition also left part of the rules. It is not difficult to see that the tree is the derivation one in the grammar $G$ if and only if there is at least one syntax cover of this tree by neighbourhoods of the defined above grammar.
\end{example}

\begin{example}{The language of structural chemical formulas}%
As it has been saying above, an each structural chemical formula may be naturally considered as a syntax diagram basing on the undirected multigraph with nodes signed by symbols of chemical elements and ribs represent covalent relations between chemical elements. Let variable $E_1$ labels any chemical element having valency $1$, $E_2$ labels any chemical element having valency $2$ and so on. Define for variable $E_1$ following neighbourhood: $\xymatrix@1{{\sf E_1} \ar@{-}[r] & E_1}$, $\xymatrix@1{{\sf E_1} \ar@{-}[r] & E_2}$ and so on for each variable $E_n$, where $n$ is the valency of a chemical element. There are two ribs in any neighbourhood of variable $E_2$. It can be either $\xymatrix@1{E_1 \ar@{-}[r] & {\sf E_2} \ar@{-}[r] & E_1}$, $\xymatrix@1{E_2 \ar@{-}[r] & {\sf E_2} \ar@{-}[r] & E_1}$, $\xymatrix@1{E_2 \ar@{-}[r] & {\sf E_2} \ar@{-}[r] & E_2}$, $\xymatrix@1{E_1 \ar@{-}[r] & {\sf E_2} \ar@{-}[r] & E_3}$ and so on, or $\xymatrix@1{{\sf E_2} \ar@{=}[r] & E_2}$, $\xymatrix@1{{\sf E_2} \ar@{=}[r] & E_3}$, $\xymatrix@1{{\sf E_2} \ar@{=}[r] & E_4}$ and etc. The same approach should be used to define neighbourhoods for variables representing elements having valencies of more high order. It is not difficult to see that the defined neighbourhood grammar describes only correct structural formulas. Let, for example, $\xymatrix@1{H \ar@{-}[r] & O \ar@{-}[r] & H}$ be the structural formula of water. There is the neighbourhood $\xymatrix@1{{\sf E_1} \ar@{-}[r] & E_2}$ for the first symbol $H$ in the diagram. For symbol $O$ it has neighbourhood $\xymatrix@1{E_1 \ar@{-}[r] & {\sf E_2} \ar@{-}[r] & E_1}$ and, for the last symbol $H$~-- neighbourhood $\xymatrix@1{{\sf E_1} \ar@{-}[r] & E_2}$. From another hand, structural formula $\xymatrix@1{H \ar@{-}[r] & H \ar@{-}[r] & H}$ is not correct in the defined neighbourhood grammar because there is not any neighbourhood, which contains the middle $H$ together with its star. The example also illustrates the need that each neighbourhood should be included to the diagram together with its star. If this not be defined there would be not any method to look at diagram $\xymatrix@1{H \ar@{-}[r] & H \ar@{-}[r] & H}$ as on incorrect one.
\end{example}

\begin{example}{The logic programming language Prolog}%
To make the idea easer to understand, will treat a Prolog-program as just a sequence of facts and rules. The fact and the rules are consisted of predicates. The predicate is a statement, which can be true or false. Syntactically the predicate contains a name and after it, in parentheses, there is a list of variables and constants where comma separate each element from an another one. The alphabet of the neighbourhood grammar for a Prolog-program  will consist of all Prolog constants. Prolog variables will be variables in the neighbourhood grammar as well. The variables as usually are the classes of symbols of alphabet (constants of Prolog-programs). A fact is the predicate, which is always true. For example, the fact $P(12,34)$ means that predicate $P$ is always true on constants $12$ and $34$. The rules of Prolog-programs have the syntax: \textit{predicate-goal} :- \textit{list of separated by commas predicate-premises}. The world of a correct Prolog-program can be described by a neighbourhood grammar in the following way. The alphabet of such the neighbourhood grammar consists of union of constants and names of predicates. The set of sorts is $S_1, S_2, \dots,S_n$ where the sort $S_i$ means the order of predicate's argument number $i$. The multigraphs of syntax diagrams are directed ones. The variables are the same as in Prolog-programs. For each constant define the neighbourhood as one symbol's diagram where the single node is signed by this constant's symbol. Such the neighbourhoods may be defined by using variables. For an each fact $P(v_1, \dots ,v_n)$, where $v_1, \ldots,v_n$ are constants and variables, define neighbourhood as the graph with nodes signed by accorded variables and constants, and ribs that connect the node signed by the predicate's name to other nodes. The sort of the rib is derived from the order of the argument of the predicate. The rib of sort $S_i$ connects the node signed by the predicate's name to the node signed by argument number $i$. The center of such the neighbourhood is the node signed by the predicate's symbol $P$. For example, for the fact $P(12,34)$ there will be the neighbourhood $\xymatrix@1{12 & P \ar@{>}[l]_1 \ar@{>}[r]^2 & 34}$. For the each rule $P(v^{P}_1, \dots ,v^{P}_n) :- P_1(v^{P_1}_1, \dots ,v^{P_1}_{n_1}), \ldots, P_k(v^{P_k}_1, \dots ,v^{P_k}_{n_k})$ define neighbourhood, which consists of nodes signed by names of predicates $P,P_1,\ldots,P_k$ and of nodes signed by names of arguments. The ribs of sorts $S_1, S_2, \dots,S_k$ connect to nodes of accorded arguments.The center of such the neighbourhood is the node signed by the name of the predicate-goal $P$. The syntax diagram, which is correct in defined above neighbourhood grammar, is one of worlds of the given Prolog-program. The elements of this world are the nodes, signed by constants and predicate names, and the ribs represent syntax relations defined by this Prolog-program. Give a little example:

\par\smallskip\noindent$\text{man}(\text{Vlad}).$
\par\smallskip\noindent$\text{man}(\text{John}).$
\par\smallskip\noindent$\text{woman}(\text{Tanya}).$
\par\smallskip\noindent$\text{pair}(X,Y):-\text{man}(X),\text{woman}(Y).$

\par\medskip\noindent
Define for this Prolog-program following neighbourhoods:
\par\smallskip\noindent$\text{\sf Vlad}$
\par\smallskip\noindent$\text{\sf John}$
\par\smallskip\noindent$\text{\sf Tanya}$
\par\smallskip\noindent$\xymatrix@1{\text{\sf man} \ar@{>}[r]^1 & \text{Vlad}}$
\par\smallskip\noindent$\xymatrix@1{\text{\sf man} \ar@{>}[r]^1 & \text{John}}$
\par\smallskip\noindent$\xymatrix@1{\text{\sf woman} \ar@{>}[r]^1 & \text{Tanya}}$
\par\smallskip\noindent
$$
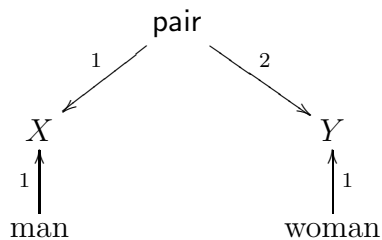
\xymatrix{
& \text{\sf pair} \ar@{>}[dl]_1 \ar@{>}[dr]^2 & \\
X & & Y \\
\text{man} \ar@{>}[u]^1 & & \text{woman} \ar@{>}[u]_1
}$$
\par\smallskip\noindent
Every neighbourhood, except the last one, is also the correct syntax diagrams in the given neighbourhood grammar. The last neighbourhood became the correct one by substituting constants instead variables $X$ and $Y$. So, the diagram 
$$
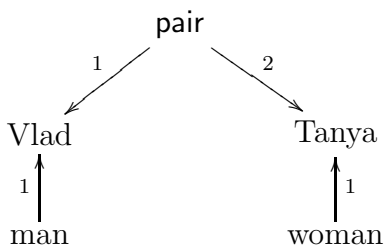
\xymatrix{
& \text{\sf pair} \ar@{>}[dl]_1 \ar@{>}[dr]^2 & \\
\text{Vlad} & & \text{Tanya} \\
\text{man} \ar@{>}[u]^1 & & \text{woman} \ar@{>}[u]_1
}$$
\par\smallskip\noindent
is the correct syntax diagram, but
$$
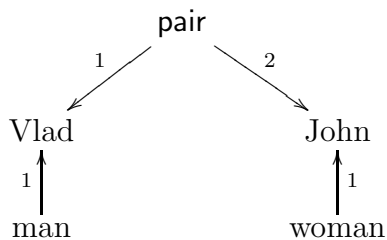
\xymatrix{
& \text{\sf pair} \ar@{>}[dl]_1 \ar@{>}[dr]^2 & \\
\text{Vlad} & & \text{John} \\
\text{man} \ar@{>}[u]^1 & & \text{woman} \ar@{>}[u]_1
}$$
\par\smallskip\noindent
is not correct because the node, signed by the symbol \text{woman}, does not include to the diagram together with any its neighbourhood.
\end{example}

\section{Conclusion}

From the author's opinion, the  given in the work approach allowing to express syntax relations of the languages by using syntax diagrams and neighbourhood grammars, is the convenient tool for formalization of languages' syntax. Especially, this covers languages with nonlinear texts. Also, in some cases, it is convenient to visualize syntax relations of the language that really exist in languages even theirs texts are linear ones. These relations are expressed implicitly in other formalisms, for example, by using rules of generative Chomsky's grammar. The approach may give the way to define some syntax properties of a language basing on their geometric representation, for example, it can be possible to define the syntactic complexity basing on some geometric properties of theirs multigraphs.

\end{document}